\documentclass{appolb}
\usepackage{epsfig}
\newcommand\beq{\begin{eqnarray}}
\newcommand\eeq{\end{eqnarray}}
\newcommand\bq{\begin{equation}}
\newcommand\eq{\end{equation}}
\newcount\eLiNe\eLiNe=\inputlineno\advance\eLiNe by -1
\begin{document}
\title{POLARIZED MUON DECAY AT REST WITH V+A INTERACTION}
\author{W. SOBK\'OW  \and S. CIECHANOWICZ 
\address{Institute of Theoretical Physics, University of Wroc\l{}aw,
Pl. M. Born 9, PL-50-204~Wroc\l{}aw, Poland\\
e-mail: {\tt ciechano@rose.ift.uni.wroc.pl}  \\ e-mail: {\tt
sobkow@rose.ift.uni.wroc.pl}}
 \and  M. Misiaszek
 \address{M. Smoluchowski
Institute of Physics, Jagiellonian University, ul. Reymonta 4,\\
PL-30-059 Krak\'ow, Poland \\
 e-mail: {\tt misiaszek@zefir.if.uj.edu.pl}}}
\maketitle
\begin{abstract}
In this paper, we analyze  the polarized muon decay at rest (PMDaR) and elastic neutrino-electron scattering (ENES) admitting the non-standard V+A interaction  in addition to standard V-A interaction.  Considerations are made for Dirac massive muon neutrino and electron antineutrino. Moreover, muon neutrinos are transversely  polarized. 
It means that the 
outgoing muon-neutrino beam is a mixture of the left- and right-chirality muon 
neutrinos and has a fixed direction of  transverse spin polarization
with respect to  production plane. 
We show that the angle-energy distribution of muon neutrinos contains the interference terms between the standard V-A and exotic V+A couplings, which are proportional to the transverse components of muon neutrino
spin polarization. They do not vanish in a limit of massless neutrino and include the relative phases to test the CP violation. In consequence, it  allows to calculate a neutrino flux and an expected event number in the ENES (detection process) both for the standard model prediction and the case of  neutrino left-right mixture.\end{abstract}


\section{Introduction}
\label{intro}
Polarized muon decay at rest (PMDaR) is the appropriate laboratory  to test
both Lorentz structure and violation of combined symmetry CP  of the  purely leptonic charged  weak interactions \cite{Jodidio,LorCP}.  In addition, the above process may be used to probe the question of lepton number violation \cite{Langacker}. The various  low energy precise measurements of  muon decay such as; the
spectral shape, angular distribution, and polarization of the outgoing electrons (positrons) led among other  to a vector-axial (V-A) structure \cite{Gell} of  the electroweak interactions of Standard Model (SM)\cite{Glashow,Wein,Salam}. This means that only left-chirality Dirac neutrinos may take
part in the charged and neutral current weak interaction. The V-A coupling also leads to maximal parity violation and massless neutrinos.   Although the SM agrees with all data up to presently available energies, it contains the large number of undetermined parameters,   so  the existence of new physics is expected. One of the fundamental aspects, which are not explained in the SM, is the origin of parity violation at current energies. In addition, the experimental precision   still does not rule out 
the participation of the exotic weak interactions with  right-chirality  neutrinos. This situation led to the formulation of various non-standard  gauge models, which have the $SU(2)_{L}\times U(1)$ as own subgroup and their predictions agree with  the SM  at low energies. Many theoretical schemes going beyond the SM include new interactions whose structure is different from V-A.   
There is a rich literature  concerning  non-standard models.   We mean here the various versions of left-right symmetric models (LRSM) with the $SU(2)_{L}\times SU(2)_{R}\times U(1)$ as the gauge group \cite{LR} and the models including mechanisms which admit scalar, tensor and pseudoscalar structures. 
The LRSM emerged first in the framework of grand unified theories (GUT).  They restore the parity symmetry at high energies and give its  violation at low energies as a result of  gauge symmetry breaking. \par 
The mentioned above muon decay is still one of the basic processes used in searching for effects connected with the $V+A$ weak interaction at low energies and possibility of CP violation.  There are many theoretical and experimental papers concerning these aspects \cite{Herczeg,Karmen,Delphi,Twist,Fetscher}. 
At present, there is no experimental evidence for the deviations  
from the SM for the Michel parameters \cite{Data}. These parameters contain only  contributions from the right-chirality neutrinos in the form of sums of the squares of certain combinations of the coupling constants.  It seems meaningful to search for new observables with the linear contributions from  the right-chirality couplings. The proper candidates can be neutrino observables, including an information on the transverse components of neutrino spin polarization, both T-even and T-odd. Today,  direct measurements of such observables in the PMDaR  are impossible, so a reasonable solution seems to be the  scattering of polarized neutrino beam coming  from the PMDaR off electrons of target. A signature of existence of the right-chirality neutrinos would be detection of  deviation from the SM prediction in the recoil electron angular distribution. 
However,  high-precision measurements of the neutrino observables require large detectors  and strong polarized (anti)neutrino sources, and long time of experiment duration (one year and longer). Moreover,  the detectors should   measure both  polar angle and azimuthal angle of the outgoing electron momentum with a high resolution, and must also distinguish the electrons from various potential background sources (e.g., the electron produced by neutrino-nucleon scattering can give a final state that is often consistent with a single recoil electron coming from neutrino-electron scattering). New types of detectors with a very low threshold $\propto 10 \,eV$ are  intensively developed by many groups. We mean the silicon cryogenic detectors based on the ionization-into-heat conversion effect and the high purity germanium detectors with the internal amplification of a signal in the electric field \cite{nmm}. It is also worth mentioning that TRIUMF Collaboration   carried out the precise measurement with the decay of polarized  muons at rest. In
order to minimize depolarization effects  the muons were
stopped in the pure metal foil and liquid-He targets \cite{Jodidio}.
\par In this paper, we focus on the study of the PMDaR (production process of polarized muon-neutrino beam) and of the ENES (detection process of exotic effects) in the presence of V+A interaction, assuming Dirac massive (anti)neutrinos. Our analysis is not made in the framework of concrete version of the left-right symmetric model. 
The main goal  is to show how the angle-energy distribution of the muon neutrinos produced in  the PMDaR depends on the interference terms between the standard vector coupling of the left-chirality neutrinos and exotic vector  couplings of the right-chirality ones. 
It will allow to calculate  the flux of muon neutrinos, both for the SM prediction and the case of neutrino left-right mixture. 
Next, we  will calculate the expected  number of events in the ENES,   when the incoming neutrino beam comes from the PMDaR and is  transversely polarized. The neutrino flux and event number are found for a detector in the shape of  flat circular ring with a low threshold.  \par  
This paper is organized as follows: Sec. II contains a basic assumptions as to the production process of muon-neutrino beam. In the sec. III,  the results for the energy-angle distribution of muon neutrinos coming from the PMDaR are presented. In the sec. IV we present the numerical results concerning the neutrino flux and expected number of events both for the SM with only left-chirality muon neutrinos and the case of left-right chirality muon-neutrino mixture,  when  muon-neutrino beam is transversely polarized. Finally, we summarize our considerations.  
\par The  
results are presented in a limit of infinitesimally small mass for all the particles produced in the muon decay. The
density operators  \cite{Michel} for the polarized initial muon and for the
polarized outgoing muon neutrino are used, see Appendix. We use the
system of natural units with $\hbar=c=1$, Dirac-Pauli representation of the
$\gamma$-matrices and the $(+, -, -, -)$ metric \cite{Greiner}.
\section{Polarized muon decay at rest - basic assumptions}
\label{sec:1}

We  assume that the polarized muon decay at rest  $(\mu^{-} \rightarrow e^- +
\overline{\nu}_{e} + \nu_{\mu})$ is a source of the muon neutrino beam. This process is described at a level of lepton-number-conserving four-fermion point interaction. 
We admit a presence of the exotic vector $g_{LR, RL, RR}^V$ 
couplings in addition to the standard vector $g_{LL}^V$ coupling. It means that
the outgoing muon neutrino flux is a mixture of the left-chirality
and  right-chirality muon neutrinos. 
 The  amplitude
for the above process is of the form:
\beq
M_{\mu^{-}} & = &
\frac{G_{F}}{\sqrt{2}}\{g_{LL}^{V}(\overline{u}_{e}\gamma_{\alpha}(1-\gamma_5)v_{\nu_{e}})
(\overline{u}_{\nu_{\mu}} \gamma^{\alpha}(1 -
\gamma_{5})u_{\mu}) \nonumber\\
&&  \mbox{} + g_{RR}^{V}(\overline{u}_{e}\gamma_{\alpha}(1+\gamma_5)v_{\nu_{e}})
(\overline{u}_{\nu_{\mu}} \gamma^{\alpha}(1 +
\gamma_{5})u_{\mu})\\
&&  \mbox{} + g_{LR}^{V}(\overline{u}_{e}\gamma_{\alpha}(1-\gamma_5)v_{\nu_{e}})
(\overline{u}_{\nu_{\mu}} \gamma^{\alpha}(1 +
\gamma_{5})u_{\mu})
\nonumber\\
&&  \mbox{} + g_{RL}^{V}(\overline{u}_{e}\gamma_{\alpha}(1+\gamma_5)v_{\nu_{e}})
(\overline{u}_{\nu_{\mu}} \gamma^{\alpha}(1 -
\gamma_{5})u_{\mu}),\nonumber
\}\eeq
where $ v_{\nu_{e}}$ and
$\overline{u}_{e}$ $(u_{\mu}\;$ and $\; \overline{u}_{\nu_{\mu}})$ are the Dirac
bispinors of the outgoing electron antineutrino and electron (initial muon and
final muon neutrino), respectively. $G_{F}= 1.16639(1)\times
10^{-5}\,\mbox{GeV}^{-2}$ \cite{Data} is the Fermi constant. The coupling
constants are denoted as $g^{V}_{LL}$  and  $g_{LR, RL, RR}^V$  respectively to
the chirality of the final electron and initial stopped muon. 
 Our analysis is carried out in the limit of massless (anti)neutrino, then 
left-chirality muon neutrino posses negative helicity, while the right-chirality one
has positive helicity, see \cite{Fetscher}.  In the
SM, only $g_{LL}^{V}$ is non-zero value. 
The table \ref{table1} displays the current limits for the $g^{V}_{LL}$, $g_{LR, RL, RR}^V$ couplings. 
\begin{table}

\begin{center}
\begin{tabular}{lll}
\hline\noalign{\smallskip}
  Coupling constants & SM & Current limits \\
  \noalign{\smallskip}\hline\noalign{\smallskip}
     $|g_{LL}^V|$ & $1$ &   $>0.960$ \\
    $ |g_{LR}^V|$ & $0$ & $<0.036$ \\
    $|g_{RL}^V|$ & $0$ & $<0.104$ \\
    $|g_{RR}^V|$ & $0$ & $<0.034$ \\
    \noalign{\smallskip}\hline
\end{tabular}
\caption{\label{table1} Current limits on the non-standard
couplings}
\end{center}
\end{table} 
Because we allow for the non-conservation of the combined symmetry CP, all the
coupling constants $g_{LL}^V, g_{LR, RL, RR}^V$   are complex.\\
The initial muon is at rest and polarized. The unit vector in the
LAB system $\mbox{\boldmath $\hat{\eta}_{\mu}$}$ denotes the muon
polarization  for a single muon decay.  
The production plane is spanned by the direction of the muon polarization
$\mbox{\boldmath $\hat{\eta}_{\mu}$}$ and of the outgoing muon neutrino
LAB momentum unit vector ${\bf \hat{q}} $. 
As is well known, in 
this plane, the polarization vector $\mbox{\boldmath $\hat{\eta}_{\mu}$}$  can
be expressed, with respect to the $\hat{\bf q}$, as a sum of the longitudinal component of the muon polarization
$(\mbox{\boldmath$\hat{\eta}_{\mu}$}\cdot\hat{\bf q}){\bf \hat{q}} $
 and transverse component of the muon polarization 
$\mbox{\boldmath $\eta_{\mu} ^{\perp}$} $, which is defined as
$ \mbox{\boldmath $\eta_{\mu}^{\perp}$} =
\mbox{\boldmath$\hat{\eta}_{\mu}$}-
(\mbox{\boldmath$\hat{\eta}_{\mu}$}\cdot\hat{\bf q}) {\bf \hat{q}}$.

\section{Energy-angle distribution of muon neutrinos}
\label{sec:2}
In this section, we show how the energy-angle distribution of  muon neutrinos  depends  
on the interference terms between the standard   and exotic
 couplings in the limit of vanishing
electron-antineutrino and muon-neutrino masses. \\
The  proper formula, obtained  after the integration over all the momentum directions of the outgoing electron and electron antineutrino,  is of the form:
\bq
   \label{didera}
\frac{d^2 \Gamma}{ dy d\Omega_\nu}
  =  
\left(\frac{d^2 \Gamma}{dy
d\Omega_\nu}\right)_{(LL+LR + RL + RR)} 
 + \left(\frac{d^2 \Gamma}{dy
d\Omega_\nu}\right)_{(INT)}
\eq 
\beq
\label{kwa}
\left(\frac{d^2 \Gamma}{ dy d\Omega_\nu}\right)_{(LL+LR + RL + RR)}   =  
\frac{G_{F}^2 m_{\mu}^5 }{768\pi^4} \bigg[ (1-\mbox{\boldmath
$\hat{\eta}_{\nu}$}\cdot\hat{\bf q}) 
(|g_{LL}^{V}|^2+|g_{RL}^{V}|^2) && \nonumber\\
 \cdot y^2( -2y+3 -(2y-1)\mbox{\boldmath $\hat{\eta}_{\mu}$}\cdot\hat{\bf q})  + (1+\mbox{\boldmath
$\hat{\eta}_{\nu}$}\cdot\hat{\bf q})(|g_{LR}^{V}|^2 +|g_{RR}^{V}|^2) && \\
\cdot y^2( -2y+3 + (2y-1)\mbox{\boldmath
$\hat{\eta}_{\mu}$}\cdot\hat{\bf q})
 \bigg],&& \nonumber\\
  \left(\frac{d^2 \Gamma}{dy d\Omega_\nu}\right)_{(INT)} = \label{INT}
\frac{G_{F}^2 m_{\mu}^5}{384\pi^4} y^2  
  \bigg[\left( Re(g_{LL}^V g_{LR}^{V*}) + Re(g_{RL}^V g_{RR}^{V*})\right)(\mbox{\boldmath
$\eta_{\nu}^{\perp}$}\cdot \mbox{\boldmath
$\hat{\eta}_{\mu}$}) &&  \nonumber \\
 - \left(Im(g_{LL}^V g_{LR}^{V*})  +  Im(g_{RL}^V g_{RR}^{V*})\right)  \mbox{\boldmath $\eta_{\nu}^{\perp}$}\cdot({\bf \hat{q}} \times
\mbox{\boldmath $\hat{\eta}_{\mu}$})  \bigg].&& 
 \eeq
Here, $y=\frac{2E_\nu}{m_{\mu}}$ is the reduced muon neutrino energy
for the muon mass $m_\mu$, it varies from $0 $ to $1$, and
$d\Omega_\nu$ is the solid angle differential for $\nu_\mu$
momentum $\hat{\bf q}$.
\\
Equation (\ref{INT}) includes the interference term between the 
$g_{LL}^{V}$ (left-chirality $\nu_\mu$) and exotic $g_{LR}^V$ (right-chirality $\nu_\mu$) couplings, so it is linear in the
exotic coupling contrary to Eq. (\ref{kwa}) and  the interference between the  $g_{RL}^V$ and  $g_{RR}^V$ couplings.  
\\
By $\mbox{\boldmath $\hat{\eta}_{\overline{\nu}}$}$,
$(\mbox{\boldmath$\hat{\eta}_{\overline{\nu}}$}\cdot\hat{\bf q}){\bf\hat{q}}$,
and $\mbox{\boldmath $\hat{\eta}_{\overline{\nu}}^{\perp}$}$ we denote the unit
polarization vector, its longitudinal component, and transverse component of the
outgoing $\nu_\mu$ in its rest system, respectively. 
\\
It is necessary to point out that there is the different dependence on the $y$ between  quadratic terms and interferences.  
For  $\mbox{\boldmath $\hat{\eta}_{\mu}$} \cdot {\bf \hat{ q}}= 0$, the
interference part can be rewritten in the following way:
 \beq\label{DDR}
 \left(\frac{d^2 \Gamma}{dy d\Omega_\nu}\right)_{(INT)} & = &
\frac{G_{F}^2 m_{\mu}^5 }{384\pi^4} |\mbox{\boldmath
$\eta_{\nu}^{\perp}$}| |\mbox{\boldmath $\eta_{\mu} ^{\perp}$}|
\\  \mbox{} \cdot \bigg\{ |g_{LL}^V ||g_{LR}^{V}|cos(\phi - \alpha)
&+&|g_{RL}^{V}||g_{RR}^{V}|cos(\phi - \beta)\bigg\}y^2, \nonumber \eeq
\\
where  $\phi$ is the angle between the direction of $\mbox{\boldmath
$\eta_{\nu}^{\perp}$}$ and the direction of $\mbox{\boldmath
$\eta_{\mu} ^{\perp}$}$ only; 
$\alpha \equiv
\alpha_{V}^{LL} - \alpha_{V}^{LR}$, $\beta \equiv
 \alpha_{V}^{RL} -
\alpha_{V}^{RR} $ are the relative phases between the $g^{V}_{LL}$,  $g^{V}_{LR}$, and   $ g^{V}_{RL}, g^{V}_{RR}$ couplings. \\
It can be noticed that the relative phases
$\alpha, \beta$ different from $0, \pi$ would indicate the CP
violation in the CC weak interaction. We also see that 
the interference terms   do not vanish in the limit of vanishing
electron-antineutrino and muon-neutrino masses. This independence of the
neutrino mass enables the measurement of the relative phases $\alpha,
\beta$ between these couplings. The interference part, Eq.
(\ref{DDR}), includes only the contributions from the transverse component of
the initial muon polarization $\mbox{\boldmath $\eta_{\mu} ^{\perp}$} $ and the
transverse component of the outgoing neutrino polarization $\mbox{\boldmath
$\eta_{\nu}^{\perp}$}$. Both transverse components are perpendicular
with respect to the $\hat{\bf q}$.
\par
Using the current data \cite{Data}, we calculate the upper limit on the
magnitude of the transverse neutrino polarization and lower bound for the
longitudinal neutrino polarization, see \cite{Fetscher}:
\beq \label{trlo}
 |\mbox{\boldmath $\eta_{\nu}^{\perp}$}|
&=& 2\sqrt{Q_{L}^{\nu}(1-Q_{L}^{\nu})}  \leq 0.103, \\
|\mbox{\boldmath $\hat{\eta}_{\nu}$}\cdot\hat{\bf q}| &=& |1- 2
Q_{L}^{\nu}|  \geq 0.995,
\\
Q_{L}^{\nu} &=& 1 - |g_{RR}^V|^{2} -  |g_{LR}^V|^{2}\geq
0.997,
\eeq
where $Q_{L}^{\nu}$ is the probability of finding  the  left-chirality $\nu_{\mu}$.
\\
If the neutrino beam comes from the unpolarized muon decay at rest, there is no interference terms in the  spectral function of muon neutrinos. \\
It is worth noticing that the effects coming from the
  neutrino mass and mixing are very small and they may be neglected. In order to show this, we use the final
density matrix for the mass states $m_1, m_2$ of $\nu_{mu}$ to avoid
breaking the fundamental principles of Quantum Field Theory. We assume that at
the neutrino detector (target)
  $\nu_\mu = cos \theta \nu_1 + sin \theta
\nu_2$. In this way, the differential neutrino spectrum is of the
form: \beq \label{mix} \frac{d^2 \Gamma}{ dy d\Omega_\nu} &=& cos^2 \theta
\frac{d^2 \Gamma}{ dy_1 d\Omega_\nu} + sin^2 \theta \frac{d^2 \Gamma}{dy_2
d\Omega_\nu}\nonumber\\ &=& \mbox{} \gamma_{(V)}^{(\phi)}\bigg[ y_{1}^2 + sin^{2} \theta \frac{\delta m_{\nu}^{2}}{m_{\mu}^2} +
O(\frac{\delta m_{\nu}^{2}}{m_{\mu}^2})\bigg], \eeq where  $\gamma_{}^{(\phi)}= \frac{G_{F}^2 m_{\mu}^5 }{384\pi^4}
    |\mbox{\boldmath $\eta_{\nu}^{\perp}$}|
|\mbox{\boldmath $\eta_{\mu} ^{\perp}$}| \bigg\{ |g_{LL}^V ||g_{LR}^{V}|cos(\phi - \alpha)$\\
$+|g_{RL}^{V}||g_{RR}^{V}|cos(\phi - \beta)\bigg\}$.  We
see that  the linear contribution from the mass mixing $\frac{\delta m_{\nu}^{2}}{m_{\mu}^2}$ is of the order of 
$10^{-19}$, taking into account the available data, so this effect  does not affect the transverse neutrino polarization. 

\section{Neutrino flux and number of events in neutrino-electron scattering}
\label{flux}
In order to find the neutrino flux, we assume that the hypothetical  detector has  the shape of flat circular ring with a low threshold  $T_{e}^{th}=10 \,eV$ $(y_{e}^{th}=0.0062)$ corresponding to  $E_{\nu}^{min}=1603.44  \, eV$. In addition, one assumes that neutrino source is located in the center of the ring detector and  polarized perpendiculary to the ring. It means that  we must integrate the angle-energy distribution of muon neutrinos over the neutrino energy in the range $[1603.44  \, eV, m_{\mu}/2]$, and over the 
$d\Omega_\nu$ in the proper range, i.e. $\phi_{\nu} \in [0,2\pi], \theta_{\nu} \in [\pi/2 -\delta, \pi/2 + \delta]$. Then, it is  multiplied  by  $\frac{N_{\mu}}{S_{D}}$ yet, where $N_{\mu}=10^{21}$ is the number of polarized muons decaying per one year.  $S_{D}=4 \pi R^2 sin \delta$, where $R=L=2205 \,cm$ is the inner radius of the detector that is equal to the distance between the muon neutrino source and detector, $\delta = 0.01$. In this way one gets the information on  the number of  muon neutrinos passing through $S_{D}$ in the direction perpendicular to the $\mbox{\boldmath $\hat{\eta}_{\mu}$}$. Fig. 1 shows the dependence of $\frac{N_{\mu}}{S_{D}} \frac{d\Gamma}{ dy} $ on the $y$ for the muon neutrinos emitted perpendicular   to $\mbox{\boldmath $\hat{\eta}_{\mu}$}$, when $V+A$ interaction is admitted. The most stringent contribution comes from the left-chirality neutrinos and it is of order of $10^{19}$. The interference gives the contribution of order of $10^{16}$.
\begin{figure}
\label{Fig1}
\begin{center}
\includegraphics*[scale=0.7]{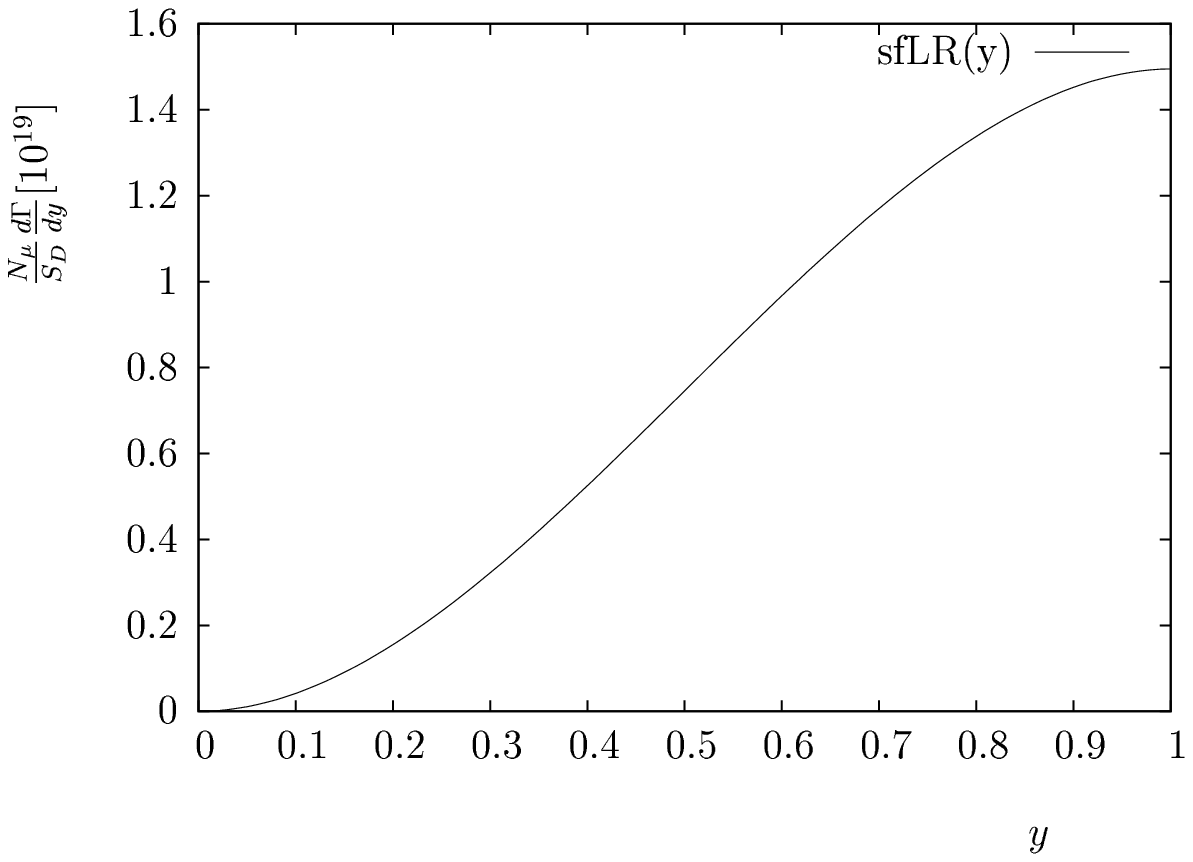}
\end{center}
\caption{Plot of the  $\frac{N_{\mu}}{S_{D}} \frac{d\Gamma}{d(cos\theta_\nu})$ as a function of $\theta_{\nu}$, when $V+A$ interaction is admitted.} 
\end{figure}
 Fig.2 illustrates the plot of $\frac{N_{\mu}}{S_{D}} \frac{d \Gamma}{d(cos\theta_{\nu})}$ as a function of $\theta_{\nu}$ (angle between   $\mbox{\boldmath $\hat{\eta}_{\mu}$}$ and $\hat{\bf q}$) for L-R mixture. The interference contribution is of order of $10^{18} sin \theta_{\nu}$. 
 \begin{figure}
\label{Fig2}
\begin{center}
\includegraphics*[scale=0.7]{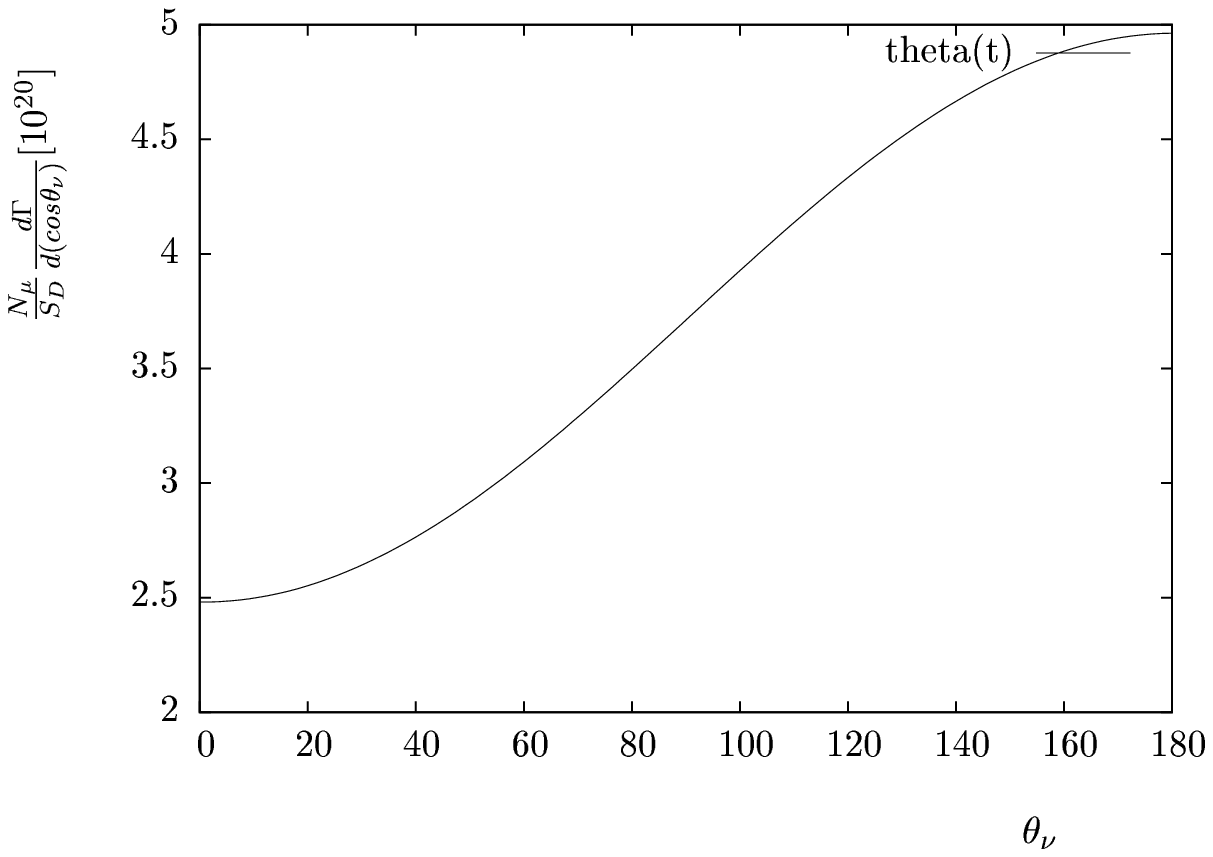}
\end{center}
\caption{Plot of the  $\frac{N_{\mu}}{S_{D}} \frac{d\Gamma}{d(cos\theta_\nu})$ as a function of $\theta_{\nu}$, when $V+A$ interaction is admitted.} 
\end{figure}
 Fig. 3 displays the dependence of neutrino flux $N_{\nu}^{\perp}$ on the $\phi$ angle for the case of CP conservation and CP violation in presence of $V+A$ interaction, when  $\hat{\bf q }\perp \mbox{\boldmath $\hat{\eta}_{\mu}$}$.  
 \begin{figure}
\label{Fig3}
\begin{center}
\includegraphics*[scale=0.7]{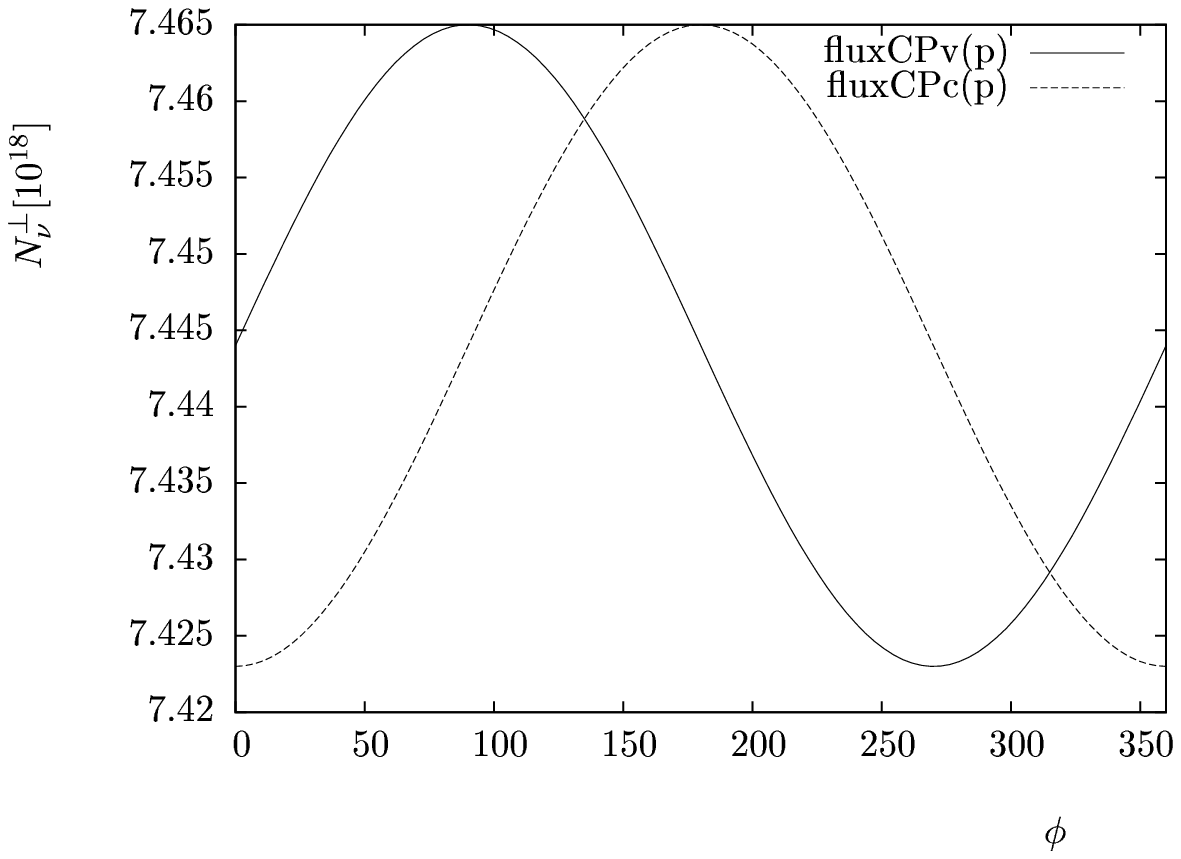}
\end{center}
\caption{Plot of the $N_{\nu}^{\perp}$ as a function  of  $\phi$: solid line shows a case of CP violation, while short-dashed line concerns  CP  conservation in presence of $V+A$ interaction, when  $\hat{\bf q} \perp \mbox{\boldmath $\hat{\eta}_{\mu}$} $. } 
\end{figure}
To calculate the expected event number,  we assume that the detector  with high efficiency ($\epsilon(y)=1$ at 
 the muon neutrino energies above threshold) contains  a large number of electrons in the fiducial volume ($N_e=2.097 \cdot 10^{34}$  corresponds to the $75$ Kton of the Fe). Moreover, the detector is able to measure both the polar angle and azimuthal angle of outgoing electrons with a high resolution.  We also need  the laboratory differential cross section for the $\nu_\mu e^-$ scattering. An appropriate   
transition amplitude includes the complex coupling constants denoted as $g_{V}^{L}, g_{A}^{L}$ and $g_{V}^{R}, g_{A}^{R}$ respectively to the initial muon neutrino of left- and right-chirality: 
\beq \label{amp} 
M_{\nu_{\mu} e} =
\frac{G_{F}}{\sqrt{2}}\{(\overline{u}_{e'}\gamma^{\alpha}(g_{V}^{L}
- g_{A}^{L}\gamma_{5})u_{e}) (\overline{u}_{\nu_{\mu'}}
\gamma_{\alpha}(1 &-& \gamma_{5})u_{\nu_{\mu}})\nonumber \\ 
  +(\overline{u}_{e'}\gamma^{\alpha}(g_{V}^{R}
+ g_{A}^{R}\gamma_{5})u_{e}) (\overline{u}_{\nu_{\mu'}}
\gamma_{\alpha}(1 + \gamma_{5})u_{\nu_{\mu}}) \},&&
 \eeq
 where $ u_{e}$ and  $\overline{u}_{e'}$
$(\overline{u}_{\nu_{\mu}}\;$ and $\; u_{\nu_{\mu'}})$ are the
Dirac bispinors of the initial and final electron (muon neutrino)
respectively;  $g_{V}^{L}= -0.035\pm  0.017, g_{A}^{L}=-0.503\pm 0.017$ \cite{Charm}; 
 \beq
g_{V}^{L}
- g_{A}^{L}\gamma_{5}&=& \frac{g_{V}^{L}
- g_{A}^{L}}{2}(1+\gamma_{5}) + \frac{g_{V}^{L}
+ g_{A}^{L}}{2}(1-\gamma_{5}),\\
g_{V}^{R}
+ g_{A}^{R}\gamma_{5}&=& \frac{g_{V}^{R}
+ g_{A}^{R}}{2}(1+\gamma_{5}) + \frac{g_{V}^{R}
- g_{A}^{R}}{2}(1-\gamma_{5}).\nonumber
\eeq 
The obtained formula has the form: 
\beq \label{przekranue} \frac{d^{2}
\sigma}{d y_{e} d \phi_{e}} = \bigg( \frac{d^{2} \sigma}{d y_{e} d
\phi_{e}}\bigg)_{(V-A)} + \bigg( \frac{d^{2} \sigma}{d y_{e} d \phi_{e}}\bigg)_{(V+A)}, && \\  
 \bigg( \frac{d^{2} \sigma}{d
y_{e} d \phi_{e}} \bigg)_{(V-A)} = B \bigg\{ (1-\mbox{\boldmath
$\hat{\eta}_{\overline{\nu}}$}\cdot\hat{\bf q}
)\bigg[(g_{V}^{L}  + g_{A}^{L})^{2}&& \\ 
+ (g_{V}^{L}- g_{A}^{L})^{2}(1-y_{e})^{2} 
 \mbox{} -  \frac{m_{e}y_{e}}{E_{\nu}}\left((g_{V}^{L})^{2} - (g_{A}^{L})^{2}\right) \bigg] \bigg\},&&
\nonumber \\ 
\bigg(\frac{d^{2} \sigma}{d y_{e} d \phi_{e}}\bigg)_{(V+A)} = B \bigg\{ (1+\mbox{\boldmath
$\hat{\eta}_{\overline{\nu}}$}\cdot\hat{\bf q}
) \bigg[(g_{V}^{R} + g_{A}^{R})^{2} && \\ + (g_{V}^{R}- g_{A}^{R})^{2}(1-y_{e})^{2}
 \mbox{} - \frac{m_{e}y_{e}}{E_{\nu}}\left((g_{V}^{R})^{2} - (g_{A}^{R})^{2}\right) \bigg] \bigg\}, && \nonumber
\eeq
where  
 $y_{e}  \equiv 
\frac{T_{e}}{E_{\nu}}=\frac{m_{e}}{E_{\nu}}\frac{2cos^{2}\theta_{e}}
{(1+\frac{m_{e}}{E_{\nu}})^{2}-cos^{2}\theta_{e}} $
is the ratio of the
kinetic energy of the recoil electron $T_{e}$  to the incoming neutrino
energy $E_{\nu}$; $B\equiv \frac{E_{\nu}m_{e}}{4\pi^2} \frac{G_{F}^{2}}{2}$;
$\theta_{e}$ is the angle between the direction of the outgoing electron momentum  $ \hat{\bf p}_{e}$  and the direction  of the incoming neutrino
momentum $\hat{\bf q}$ (recoil electron scattering angle); $m_{e}$ is the electron mass; 
$\phi_{e}$ is the 
angle between the production plane and the reaction plane spanned by the $
\hat{\bf p}_{e}$ and $ \hat{\bf q}$.\\ The above formula  describes the scattering of transversely polarized muon-neutrino   beam on the unpolarized electrons in the limit of vanishing neutrino mass. In our case the  incoming neutrino beam is the mixture of the 
left-chirality  neutrinos detected in the standard $g_{V}^{L}, g_{A}^{L}$ weak interactions and right-chirality  ones detected in the exotic $g_{V}^{R}, g_{A}^{R}$ weak interactions. 
 We see that all the interference terms between the
standard and exotic couplings vanish in the massless neutrino limit. There are only squared contributions from exotic couplings proportional to the the longitudinal  neutrino polarization. 
 It is necessary to point out that there is an observable  containing the linear contributions from the exotic couplings which are independent of the neutrino mass. We mean the angular distribution of outgoing neutrinos, but such a measurement is totally  unrealistic.  
\\
 The number of events is found as result of integration of the differential cross section over the $T_{e}$ in the range $[T_{e}^{th}=10 \,eV, T_{e}(E_{\nu})]$. So, the received expression together with the spectral function  integrates  over the $E_\nu$ in the range $[1603.44  \, eV, m_{\mu}/2 )]$. Finally, we must multiply it by $N_{\mu}/S_{D}$.   
\par Using the available data \cite{Data, Charm, Rod}, we get the flux of neutrino beam and the number of events  both for the SM and the case of left-right mixture, see table \ref{table2}.  
\begin{table}
\begin{tabular}{lll}
\hline\noalign{\smallskip}
 Case & Neutrino flux  & Event number\\
  & $ \Phi_{\overline{\nu}}^{\perp} [cm^{-2}s^{-1}] $ & $ N_{e}$\\
  \noalign{\smallskip}\hline\noalign{\smallskip}
 1. SM & $7.483 \cdot 10^{18} $ &   $8.854 \cdot 10^9$ \\
  $2. g_{LL}^{V} + g_{RL}^{V}+g_{LR}^{V} + g_{RR}^{V}$  & $7.444 \cdot 10^{18}$ & $9.854\cdot 10^9$ \\
$3. g_{LL}^{V}g_{LR}^{V}+g_{RL}^{V}g_{RR}^{V}$ & $2.079 \cdot 10^{16}$ & $ 2.95\cdot 10^7 $  \\ 
$4. 2+3 $ & $7.465 \cdot 10^{18}$ & $9.884 \cdot 10^9$ \\
 \noalign{\smallskip}\hline
\end{tabular}
\caption{\label{table2} The  values of neutrino flux and event number predicted  per one year for the SM, and in  the presence of $V+A$ interaction.}    
\end{table}
\section{Conclusions}
\label{sec:5}
In this paper, we investigated the PMDaR and ENES in the presence of the exotic $V+A$ interaction, when the (anti)neutrinos have Dirac nature and are massive.\\ We have shown that the angle-energy distribution of muon neutrinos produced in the PMDaR includes the terms with interference between the $g_{LL}^{V}$ (left-chirality $\nu_\mu$) and exotic $g_{LR}^V$ (right-chirality $\nu_\mu$) couplings, which are independent of the muon neutrino and electron antineutrino masses. These interferences are exclusively   proportional to the T-even and T-odd  transverse components of neutrino polarization $\mbox{\boldmath $\eta_{\mu} ^{\perp}$}$. \\
Next, we have calculated the angular distribution of recoil electrons for the ENES, when the incoming polarized neutrino beam comes from the PMDaR. We have demonstrated that the laboratory differential   cross section contains only the squared contributions from the exotic couplings, because all the interferences are suppressed by the neutrino mass. It means that the angular distribution should be azimuthally symmetric. It is worth noting that the SM also predicts  the similar regularity for the angular distribution.\\
Using the current data, we have computed the flux of muon  neutrinos and expected event number for a given configuration of detector, both for the SM prediction and for the case of L-R neutrino mixture. \\
 Unfortunately, the detection of new effects at  present experimental precision would be very difficult, even if  one had the strong neutrino source  ($10^{21}$ or more polarized  muons decaying at rest per year) and the  high-precision large detector ($10^{34}$ or more target-electrons) with the low threshold, measuring  both  polar angle and azimuthal angle of the outgoing electron momentum. 
\\We have also displayed that the eventual effects connected with the neutrino mass and mixing in the spectral function are totally inessential $( \sim 10^{-19})$.  \\
We plan to search for the other polarized (anti)neutrino beams, which could be
interesting from the aspect of observable effects caused by the exotic
right-chirality states. We expect  some interest of the neutrino laboratories working with polarized muon decay and artificial  (anti)neutrino sources, and neutrino beams, e.g. KARMEN, PSI, TRIUMF,
BooNE, SUPERKAMIOKANDE. 

\appendix 
\section{Appendix: Four-vector of antineutrino spin polarization and density operator}
\label{app1}
The formula for the  spin polarization 4-vector of  massive neutrino
$S^\prime$
moving  with the momentum ${\bf q}$ is as follows:
\beq
S^\prime & = & (S^{\prime 0}, {\bf S^\prime}),\\
S^{\prime 0} & = & \frac{{|\bf q|}}{m_{\nu}}(\mbox{\boldmath
$\hat{\eta}_{\nu}$}\cdot{\bf \hat{q}}),
\\
{\bf S^\prime} & = &
\left(\frac{E_{\nu}}{m_{\nu}}(\mbox{\boldmath
$\hat{\eta}_{\nu}$}\cdot{\bf \hat{q}}){\bf \hat{q}} +
\mbox{\boldmath $\hat{\eta}_{\nu}$} - (\mbox{\boldmath
$\hat{\eta}_{\nu}$}\cdot{\bf \hat{q}}){\bf \hat{q}}\right),
\eeq
where $\mbox{\boldmath $\hat{\eta}_{\nu}$}$ is the unit 3-vector of
the neutrino polarization in its rest frame. The formula for the density
operator of the polarized neutrino in the limit of vanishing  neutrino
mass $m_{\nu} $  is given by:
\beq
\lim_{m_{\nu}\rightarrow 0}\Lambda_{\nu}^{(s)} &=&
\lim_{m_{\nu}\rightarrow 0}
\frac{1}{2}\bigg\{\left[(q^{\mu}\gamma_{\mu}) + m_{\nu}\right]\left[1 +
\gamma_{5}(S^{\prime \mu}\gamma_{\mu})\right]\bigg\}\nonumber  \\
  =    \frac{1}{2}&\bigg\{&(q^{\mu}\gamma_{\mu})
 \left[1 + \gamma_{5}(\mbox{\boldmath
$\hat{\eta}_{\nu}$}\cdot{\bf \hat{q}}) + \gamma_{5}
S^{\prime\perp}\cdot \gamma \right]\bigg\},
\eeq
where $S^{\prime\perp} =
\left(0, \mbox{\boldmath $\eta_{\nu} ^{\perp}$} =  \mbox{\boldmath
$\hat{\eta}_{\nu}$} - (\mbox{\boldmath
$\hat{\eta}_{\nu}$}\cdot{\bf \hat{q}}){\bf \hat{q}}\right)$. We see
that in spite of the singularity $m_{\nu}^{-1}$ in the polarization four-vector
$S^\prime $, the density operator $\Lambda_{\nu}^{(s)}$ 
including the transverse component of the neutrino spin polarization
remains finite \cite{Michel}.


\begin{thebibliography}{99}
%
\bibitem{Jodidio} A. Jodidio et al., {\sl Phys. Rev.} {\bf D 34},  1967 (1986).
\bibitem{LorCP} K. Mursula, M. Ross, F. Scheck, {\sl Nucl. Phys.} {\bf B 219}, 321 (1983); J. Maalampi, K. Mursula, M. Ross, {\sl Nucl. Phys.} {\bf B 207}, 233 (1982). 
 \bibitem{Langacker} P. Langacker, D. London, {\sl Phys. Rev.} {\bf D 39}, 266 (1989). 
\bibitem{Gell} R. P. Feynman, M. Gell-Mann, {\sl Phys. Rev.} {\bf 109},    193 (1958);
  E. C. G. Sudarshan, R. E. Marshak, {\sl Phys. Rev.}  {\bf 109},   1860 (1958).
\bibitem{Glashow} S. L. Glashow,  {\sl Nucl. Phys.}   {\bf 22},  579 (1961).
\bibitem{Wein} S. Weinberg,  {\sl Phys. Rev. Lett.}  {\bf 19},    1264 (1967).
\bibitem{Salam} A. Salam, in(Ed.),
 Elementary Particle Theory, ed. N. Svartholm  (Almqvist and Wiksell, Stockholm, 1968), 367.
 \bibitem{LR} J. C. Pati, A. Salam, {\sl Phys. Rev.} {\bf D 10}, 275 (1974); R. N. Mohapatra, J. C. Pati,  {\sl Phys. Rev.} {\bf D 11}, 2558 (1975); R. N. Mohapatra, G. Senjanovic, {\sl Phys. Rev.} {\bf D 12}, 1502 (1975); {\sl Phys. Rev. Lett.} {\bf 44}, 912 (1980);   {\sl Phys. Rev.} {\bf D 23}, 165 (1981); {\sl Nucl. Phys.} {B 153}, 334 (1979); R. N. Mohapatra, {\sl Prog. Part. Nucl. Phys. } {\bf 26}, 1  (1992). 
\bibitem{Herczeg} P. Herczeg, {\sl Phys. Rev.} {\bf D 34}, 3449 (1986).
\bibitem{Karmen} B. Armbruster et al., {\sl Phys. Rev. Lett.} {\bf 81}, 520 (1998).
\bibitem{Delphi}  DELPHI Collaboration, P. Abreu  et al.,  {\sl Eur. Phys. J.}   {\bf C 16},   229 (2000).
\bibitem{Twist} TWIST Collaboration, J. R. Musser et al., {\sl Phys. Rev. Lett.} {\bf 94},  101805 (2005).
\bibitem{Fetscher} W. Fetscher, {Phys. Rev.} {\bf D 49},  5945 (1994).
\bibitem{Data}  C. Amsler et al., {\sl Phys.Lett.} {\bf B 667}, 1 (2008).
\bibitem{nmm} B.S. Neganov et al., hep-ex/0105083. 
\bibitem{Michel} L. Michel, A. S. Wightman,  {\sl Phys. Rev.
} {\bf 98},   1190 (1955).
\bibitem{Greiner} W. Greiner, B. Muller, Gauge Theory of Weak Interactions, Springer, 2000.
\bibitem{Charm} CHARM
II Collaboration, {\sl Phys. Lett.} {\bf B 332}, 465 (1994); {\sl Phys. Lett.} {\bf B 345}, 115 (1995).
\bibitem{Rod} A. Gutierrez-Rodriguez, M. A. Hernandez-Ruiz, M. Maya, hep-ph/0111011. 
\end{thebibliography}
\end{document}